\documentstyle[sprocl]{article}

 \catcode`@=11
\def\@cite#1#2{\unskip\nobreak\relax
    \def\@tempa{$\m@th^{\hbox{\the\scriptfont0 #1}}$}%
    \futurelet\@tempc\@citexx}
\def\@citexx{\ifx.\@tempc\let\@tempd=\@citepunct\else
    \ifx,\@tempc\let\@tempd=\@citepunct\else
    \ifx;\@tempc\let\@tempd=\@citepunct\else
    \ifx:\@tempc\let\@tempd=\@citepunct\else
\let\@tempd=\@tempa\fi\fi\fi\fi\@tempd}
\def\@citepunct{\@tempc\edef\@sf{\spacefactor=\the\spacefactor\relax}\@tempa
    \@sf\@gobble}

\def\lsim{\mathrel{\raise.2ex\hbox{$<$}\hskip-.8em\lower.9ex\hbox{$\sim$}}}
\def\gsim{\mathrel{\raise.2ex\hbox{$>$}\hskip-.8em\lower.9ex\hbox{$\sim$}}}

\begin{document}
\thispagestyle{empty}

\vspace*{-.7in}

\font\fortssbx=cmssbx10 scaled \magstep1
\hbox to \hsize{
%\special{psfile=uwlogo.ps hscale=7000 vscale=7000 hoffset=-12 voffset=-2}
%\hskip.35in \raise.1in
\hbox{\fortssbx University of Wisconsin - Madison}
\hfill$\vcenter{\hbox{\bf MADPH-99-1099}
                \hbox{January 1999}}$ }

\bigskip

\title{\uppercase{New Worlds in Astroparticle Physics:\\
 Summary Talk}\footnote{Talk given at {\it Second Meeting on ``New Worlds in
Astroparticle Physics"}, University of the Algarve, Faro, Portugal (1998).}
}

\author{\unskip\medskip\uppercase{Francis~Halzen}}
\address{Department of Physics, University of Wisconsin, Madison, WI 53706}

\maketitle
\abstracts{
Surrounded by stunning Algarve landscapes not far from where Henry the
Navigator organized the voyages that mapped the Earth, particle astrophysicists
discussed new initiatives to explore the cosmos. While first generation
experiments opened new voyages of the mind with evidence for neutrino mass and
a cosmological constant, much of the discussion focussed on novel experimental
assaults on the secrets of the Universe. While ``big-time" particle physics
entered space with AMS and high energy neutrino telescopes saw first light,
neutrinos actually, at Lake Baikal and South Pole, it is the hope that new and
even more ambitious experimental initiatives, ranging from gravitational wave
detectors to the MAP and Planck microwave satellite-borne detectors, will boost
particle astrophysics into ``the" physics and astronomy of the next
millennium.}

\section[]{Introduction}

Efforts are underway to qualitatively improve the instruments that can push
astronomy beyond GeV photon energy, to wavelengths smaller than $10^{-14}$\,cm,
and map the sky in neutrinos and EeV cosmic rays as well as gamma rays. New
gravitational wave detectors will explore wavelengths much larger than those of
radio astronomy. While particle astrophysics may be easily mistaken for
astronomy, I notice that most participants at this meeting are card-carrying
particle physicists, born and raised near accelerators. Particle astrophysics
presents particle physics with extraordinary opportunities. With neutrino mass,
the cosmological constant and dark matter as some of the topics dominating this
meeting, the case is self-evident.

Particle physics forms a basic framework which has allowed us to launch some of
the most far-reaching excursions of the mind into the structure of matter and
of the Universe. Further progress will come both from pushing the high energy
and high sensitivity limits at accelerators\cite{accelerator}, and from
vigorously exploring the interfaces with other fields. Particle astrophysics
has been one of the more successful of these multidisciplinary ventures.

The symbiosis of particle physics and astrophysics is even more intimate when
it comes to instrumentation. The construction of dark matter detectors, earth-
and space-based gamma ray telescopes, giant natural neutrino detectors and
state-of-the-art air shower detectors, is immersed in technology developed for
accelerator experiments. Particle physicists should not be reluctant in
entering these new interdisciplinary ventures which touch astronomy,
astrophysics and cosmic ray physics using instrumentation that is a direct
spin-off from techniques developed at great effort and expense in our
accelerator laboratories. Second generation particle astrophysics experiments
will require frontier technology, even by particle physics
standards\cite{gilman}.

Cosmic beams of photons and protons have been detected with energies far
exceeding those within reach of accelerators. How and where Nature accelerates
particles to these energies is still a matter of speculation. We have learned
that the sun does indeed emit a few percent of its energy in neutrinos. What
were at first routine studies of cosmic ray interactions in the atmosphere,
produced indications that neutrinos have mass. Supernova 1987A told us that we
do understand stellar collapse, and also delivered a limit on the mass of the
electron-neutrino similar to laboratory experiments. Its observation severely
constrained the mass of the axion. Atmospheric Cherenkov telescopes have
unambiguously detected several sources in gamma rays of energies between 1 and
10 TeV and are now providing a new window on the most violent sites in the
cosmos. Finally, novel techniques developed by particle physicists at Berkeley
produced evidence for a cosmological constant in observations of high red-shift
supernovae.

Future goals hold promise beyond past achievements with, for instance, the
possibility of detecting the particles which constitute the dark matter. The
apparent relationship between the electroweak scale and the mass density of a
flat universe is one of the most intriguing hints in contemporary science. The
search for the particle nature of dark matter and the study of high energy
neutrinos are examples of many common intellectual endeavors of particle
physics and particle astrophysics. Also, the large neutrino detectors which
have yielded tantalizing hints of new physics in the atmospheric neutrino beam,
possibly neutrino mass, are complemented by new, giant detectors which are,
hopefully, large enough to study neutrino sources far beyond our own galaxy.
Gauge theory, which is the basic framework of modern particle physics, suggests
topological structures which can only be probed in non-accelerator experiments.
Such topological defects may manifest themselves both in cosmology (for
instance, generating structure in the cosmic microwave background and the large
scale distribution of galaxies) and in the acceleration of the highest energy
cosmic rays.

Doing particle physics beyond the boundaries of the accelerator laboratories,
for instance in space, at the South Pole or in the deep ocean, will inspire
future generations of scientists and the public at large\cite{gilman}.

I will briefly summarize the excursions into particle astrophysics emphasized
at this meeting:

\begin{enumerate}
\addtolength{\itemsep}{-2mm}
\item
{\bf neutrinos}: first evidence for oscillations, and first light for
first-generation neutrino telescopes,

\item
{\bf gamma rays}: first light for next generation ground-based detectors using
solar power stations and for the MILAGRO detector,

\item
{\bf EeV-protons}: the particles that do exist, but shouldn't,

\item
{\bf cosmology and gravity}: $\Omega = 1$, but $\Omega_{matter} \simeq 0.4$ ---
the cosmological constant, eighty years later,

\item
{\bf dark matter}: the particles that should exist, but don't.

\end{enumerate}

It would be unwise however for the uninitiated observer to try to summarize the
imaginative and sometimes heroic incursions of the theorists into the
astrophysics of neutrinos. I especially enjoyed the discussions of neutrino
magnetic moments\cite{mu}, and of the collective interactions of neutrinos in
astronomical plasmas\cite{collective}.

\section[]{Neutrinos\cite{neutrino}}

Neutrino astronomy was born with the sun in a series of pioneering efforts
starting with the first observation in the Homestake mine and culminating with
the GALLEX and SAGE experiments which detected the dominant solar source of
neutrino production by proton-proton fusion. With the present emphasis on (and
sometimes controversy surrounding) ``the deficit", the primary achievement of
these historic experiments to see the sun in neutrinos should not be forgotten.
Consolidation of the speculations that the solar deficit indicates an
non-vanishing mass of the $\nu_e$ requires help from experiment which may very
well come from SuperKamiokande. Supporting evidence may come as a distortion of
the neutrino energy spectrum, or a daily or seasonal variation of the flux. The
case for a particle, rather than astrophysical solution, has been boosted by
ever more precise helioseismology. A lineup of new solar neutrino detectors
lead by SNO and Borexino is ready to tackle the problem. A second-generation
GALLEX experiment was discussed at this meeting.

While many now feel that neutrino mass has been finally established, the
evidence came from elsewhere. Pathological behavior of the neutrinos produced
in cosmic ray interactions with atmospheric nuclei, has been established for
some time. The observed ratio of neutrinos of electron and muon type in the
atmospheric beam disagrees with a very solid theoretical prediction. While this
discrepancy can be readily accommodated by assuming oscillation of the neutrino
beam, first support for this interpretation came from the SuperKamiokande
experiment with a most striking and straightforward observation: there are
fewer muon neutrinos produced in the earth's atmosphere below our feet than
above their head, a reduction in relative flux of more than 6~$\sigma$. So, the
anomalous ratio of electron and muon neutrinos can be traced to a reduction of
the muon neutrino flux which travelled 12500~km from the other end of the
earth, relative to the flux produced in the upper atmosphere overhead which
travelled, on average, 25~km. Supporting observations by SuperK and other
experiments confirm this interpretation; none are however compelling.

The data is described by a mixing angle near unity and a mass difference
$\Delta m^2$ of a few times $10^{-3}$~$eV^2$. What now? Clearly some, waiting
for decades for a crack in the harness of the Standard Model, have already
built houses of cards. Even by the most conservative interpretation, this
result must have truly fundamental implications. One can accommodate the mass
by adding a right-handed singlet to the Standard Model fermion multiplet as in
its SO(10) extension. A mass term is added to the Lagrangian which represents
new physics with a coupling $\lambda$ given by
\begin{equation}
m_{\nu} = \lambda v^2 \simeq 0.1\rm\ eV^2.
\end{equation}
With a Standard Model vacuum expectation value of 250~GeV, this calls for new
physics at an energy scale $M$
\begin{equation}
\lambda = M^{-1} = {6 \times 10^{14}\rm\, GeV}^{-1},
\end{equation}
possibly larger, and not too far from the Planck scale. With sub-eV masses,
neutrinos are not dark matter, mixed or not. Remember however that even a
conservative house of cards is a house of cards and that the experiments
measure $\Delta m^2$ and not $m$.

Unusually fundamental results require confirmation of unusual quality:
possibly, observing the reappearance of the $\nu_{\mu}$ beam as $\nu_{\tau}$'s.
Long-baseline experiments provide the best hope, although present results call
for a baseline in excess of 1000~km which none of the present proposals
deliver. This number does have a large uncertainty. If everything else fails,
one may have to look elsewhere, for instance lowering the threshold of high
energy neutrino telescopes.

This speculation is no longer unrealistic: first neutrino events emerged from
Lake Baikal water and South Pole ice. Construction and calibration of their
first-generation instruments was completed in the months preceding this
meeting. First calibration of the respective experimental techniques using the
atmospheric neutrino beam is now possible. There already are immediate
implications beyond the obvious: relatively shallow experiments can handle the
large cosmic ray muon backgrounds, and one can reconstruct tracks in ice
opening the possibility of commissioning a kilometer-scale detector in the near
future. A wide variety of estimates indicate that this is the size of detector
required to do the science. Consisting of several hundred optical modules
deployed in natural water or ice which acts as a Cherenkov medium, these
detectors are optimized for large effective area rather than low threshold
(10~GeV or higher, even for the present smaller versions).

These instruments are complementary to SuperK and exploiting them to confirm
their atmospheric neutrino results will be challenging. Lowering the threshold
by redesigning the telescope architecture is not, and can be achieved by
reducing the spacings of the optical modules in all, or part of the detector.
Doing this may not further their astronomical mission, but will turn these
instruments into better atmospheric neutrino detectors, good enough to probe
the SuperK signatures for neutrino mass. The South Pole experiment would also
have the right baseline to receive an accelerator beam.

Several initiatives exist to develop the infrastructure and technologies for
the deployment of a neutrino telescope in the Mediterranean basin. At this
meeting the Antares collaboration revealed, after satisfactory initial tests,
their plans to proceed with the deployment of a first string of optical modules
in 99, the construction of a detector of 800 modules on 10$\sim$15~strings by
02, and a kilometer-scale detector by 06.

As with conventional telescopes, at least two are required to cover the sky. As
with particle physics collider experiments, it is very advantageous to explore
a new frontier with two or more instruments, preferably using different
techniques. This goal may be achieved by exploiting the parallel efforts to use
natural water and ice as the Cherenkov medium for particle detection. Let me
conclude by trying to infuse some sanity in the non-debate on ``water and ice".
It is a non-debate because, ideally, we want both. Given the pioneering and
exploratory nature of the research, we most likely {\bf need} both. Water and
ice have complementary optical properties: while the ``attenuation" lengths are
comparable for the blue wavelength photons relevant to the experiments,
attenuation is dominated by scattering in ice and by absorption in water. Both
have a problem: scattering in ice, potassium decay and bioluminescence in
water. Both problems can be solved as shown by the initial results.

\section[]{Gamma Ray Astronomy on Earth and in Space\protect\cite{gamma}}

State-of-the-art particle physics technology has reached space with the AMS
anti-matter spectrometer which made a successful flight on the NASA shuttle.
The field of gamma ray astronomy is buzzing with activity to construct
second-generation instruments. Space-based detectors are extending their reach
from GeV to TeV energy with AMS and, especially, GLAST, while the ground-based
Cherenkov telescopes are designing instruments with lower thresholds. In the
not so far future both techniques should generate overlapping measurements in
the $10{\sim}10^2$~GeV energy range. All ground-based experiments reach for
lower threshold, better angular- and energy-resolution, and a longer duty
cycle. One can identify a multi-prong attack, with different methods for
improving air Cherenkov telescopes:

\renewcommand{\theenumi}{{\it\roman{enumi}}}
\begin{enumerate}

\item
larger mirror area, exploiting the parasitic use of solar collectors during
nighttime,
\item
better, or rather, ultimate imaging of the photon footprint in the atmosphere
with the 17~m MAGIC mirror,
\item
larger field of view by using multiple telescopes.

\end{enumerate}

At this conference the first results from the CELESTE instrument were reported.
Atmospheric air showers initiated by photons are imaged using an abandoned
solar power station in the French Pyrenees. Each heliostat is viewed by a
photomultiplier via optics placed at the focus, in the tower where solar power
was once harnessed. The technique has been demonstrated by observing the Crab
supernova remnant with a threshold of 80~GeV using only 18 heliostats and 9
data acquisition channels triggering at 10~Hz. This threshold corresponds to
only 4 photons per heliostat. The march to lower threshold is on track.

After two decades, ground-based gamma ray astronomy has become a mature
science. Let me remind you that, although it has produced few sources by
astronomical standards, their observation has produced spectacular results.
Data taken on the flaring active galaxies Markarian 421 and 501 testify to this
statement. The most prominent features are:

\begin{itemize}

\item
a spectrum which extends beyond 30~TeV,
\item
emission of TeV-photons in bursts with a duration of order a few days,
\item
correlation between the optical and TeV variability,
\item
observation of a burst lasting only 15 minutes, suggesting emission from very
localized regions of the galaxy, presumably the jet.

\end{itemize}

There is a dark horse in this race: Milagro. The Milagro idea is to lower the
threshold of conventional air shower arrays to 100~GeV by uniformly
instrumenting an area of $10^3$~m$^2$ or more (no sampling!). For time-varying
signals, such as bursts, the threshold could be even lower. One instruments a
pond with photomultipliers (Milagro), or covers a large area with resistive
plate chambers (ARGO), or even with muon detectors (Hanul) which identify point
sources of muons produced in photon-induced air showers.

\section[]{Proton Astronomy: EeV Cosmic Rays\protect\cite{cosmic}}

Around 1930 Rossi and collaborators discovered that the bulk of the cosmic
radiation is not made up of gamma rays. This marked the beginning of what was
then called ``the new astronomy'', and we refer to as cosmic ray physics today.
It is ``astronomy" only above $5 \times 10^{19}$~eV or so, where the arrival
directions of the charged cosmic rays are not scrambled by the ambient magnetic
field of our own galaxy. We suspect that the bulk of the cosmic rays are
accelerated in the blastwaves of supernovae exploding into the interstellar
medium. This mechanism has the potential to accelerate particles up to energies
of $10^3$~TeV where the cosmic ray spectrum suddenly steepens: the ``knee" in
the energy spectrum. We have no clue where and how cosmic rays with energies in
excess of $10^3$~TeV are accelerated. We are not even sure whether they are
protons or iron, or anything else. The origin of cosmic rays with energy beyond
the ``knee" is one of the oldest unresolved puzzles in science.

To illustrate the degree of desperation, it has been suggested that the highest
energy cosmic rays are the decay products of $10^{24}$~eV (the GUT unification
scale) topological defects such as a monopoles, strings\dots. Topological
structures are deeply connected to gauge theories and cannot be studied in
accelerator experiments. Non-accelerator particle physics provides unique
opportunities here. A topological defect will suffer a chain decay into GUT
particles X,Y, which subsequently decay to the familiar weak bosons, leptons
and quark-gluon jets. Cosmic ray protons are the fragmentation products of
these jets.

If the sources of cosmic rays are beyond $10^2$~Mpc, conventional astronomy
cannot identify them because of the absorption of the beam on the microwave
background. Absence of an energy cutoff associated with this absorption (the
Greissen-Kuzmin-Zatsepin cutoff) becomes a signature for distant sources. The
main problem today is statistics. After particles with energies in the vicinity
of 100~EeV were discovered at Haverah Park, we have accumulated some 10 events
whose energy clearly exceed $10^{20}$~eV, using three different detectors:
AGASA, Yakutsk and the Fly's Eye. The latter is being replaced by a technically
superior instrument with larger collection area: the HIRES detector.
Construction of a $10^4$~km$^2$ array, one hundred times larger than the AGASA
array operating in Japan, has been proposed and will be launched soon as the
``Auger" project.

\section[]{Gravity\cite{gravity}, Cosmology\cite{cosmology} and Dark
Matter\cite{dark}}

The asymmetric collapse, e.g.\ of a rotating star, near the center of our
galaxy will result in the supernova display astronomy is waiting for --- the
simultaneous observation of light, neutrinos and gravitational waves could be
the scientific event of all times. If we make the optimistic assumption that a
similar amount of energy is emitted in gravitational waves and in light, i.e.\
one hundredth of a solar mass, the new generation of gravitational antennas
under construction in the US and Europe will detect a whopping signal of
$\delta h=10^{-18}$. This deformation of the transverse components of the
space-time tensor $h^{TT}_{\mu\nu}(x-ct)$ is detected at Earth in the form of
gravitational waves. Such endeavors have put general relativity back into the
particle physicist's bag of tools. We were reminded at this meeting that this,
and other adventures involving the cosmological constant and dark matter, are
built on a total faith of the framework.

Concerning gravity, theorists have been investigating the interesting
suggestion that the Planck scale is of order the weak scale of 1~TeV. At the
Planck, the other particle interactions cannot be separated from gravity: a
particles's Compton wavelength ($m^{-1}$) is of the same order of magnitude as
its Schwarzschild radius ($G_Nm$). The idea implies that particle interactions
modify gravity for distances below 1 millimeter! It is amazing to realize that
no experimental verifications of Newtonian gravity cover this regime (yet), and
the Large Hadron Collider will study gravity.

Even with a Planck scale safely anchored at $10^{19}$~GeV, cosmology has become
(too?) exciting. The convergence on a Standard Cosmology with a flat Universe
with $\Omega =1$ has been shattered by multiple blows: we now suspect that
$\Omega \leq1$ for matter and that the cosmological constant does not vanish.
The latter is an awesome possibility. While we know the particle physics that
rules the other great epochs of cosmology, nucleosynthesis and recombination,
we do not know the physics that rules the expanding Universe we live in today,
driven by a cosmological constant.

Although the evidence, presented here by two groups, emerged from measurements
of the Hubble flow using supernovae as standard candles, corroborating evidence
may be emerging from elsewhere. Recent South Pole measurements\cite{texas} of
the acoustic waves at the surface of last scattering (the so-called Doppler
peak in the power spectrum), indicate that $\Omega =1$. With $\Omega_{matter}
\leq 1$, this leaves room for a cosmological constant closing the deficit.

We were reminded at this conference that the search for particle dark matter is
still a main focus of particle physicists entering astroparticle physics. This
search is reaching the critical point where the size and sensitivity of the
experiments will reach the predictions of the most popular model: neutralino
dark matter made of the lightest stable particle predicted by supersymmetry.
Phonon, scintillation and other techniques are developed, often in experiments
exploiting coincident signals. Where these experiment lose sensitivity with
increasing neutralino mass, the now-operating neutrino telescopes gain
sensitivity all the way to TeV masses, the maximum allowed by Standard
Cosmology. {\bf High} mass neutralinos annihilate in sun and earth into {\bf
high} energy neutrinos which are easier to detect.

We feel that in astroparticle physics, like in astronomy, mother Nature is
always more imaginative than scientists. The future of astroparticle physics is
not only bright --- I predict, with history on my side, that it will be
brighter than we can imagine.

\section*{Acknowledgements}

The hospitality of Jorge Dias de Deus, his colleagues at the University of the
Algarve and our many friends at other Portuguese universities, has become
legendary in only two meetings. I have the feeling that this series of
conferences has as bright a future as its subject. Thanks.

This work was supported in part by the University of Wisconsin Research
Committee with funds granted by the Wisconsin Alumni Research Foundation, and
in part by the U.S.~Department of Energy under Grant No.~DE-FG02-95ER40896.

\end{document}